\documentclass[12pt]{article}
\usepackage[T2A]{fontenc}
\usepackage[cp1251]{inputenc}
\usepackage[english]{babel}
\usepackage[dvips]{graphicx}
\normalsize
\usepackage{amssymb,amsmath}

\sloppypar

\def\vt{$\xi_{\rm t}$}

\newcommand{\se}{$S_{\rm e^-}$}

\newcommand{\eps}{\log\varepsilon}
\newcommand{\kms}{km\,c$^{-1}$}

\newcommand{\kH}{$S_{\rm H}$}    

\begin{document}
\selectlanguage{english} 

\title{Influence of Departures from LTE on Oxygen Abundance Determination}

\author{T. ~M. ~Sitnova \thanks{sitnova@inasan.ru}\;\;$^{1,2}$\; L. ~I.~Mashonkina$^1$, T. ~A. ~Ryabchikova$^1$\\[2mm]
\begin{tabular}{l}
 $^1$ {\it Institut of Astronomy of RAS, Moscow, Russia}\\[2mm]
 $^2$ {\it Moscow M.V. Lomonosov State University, Moscow, 119991 Russia}
\end{tabular}
}
\date{}
\maketitle
 





\begin{abstract}

We have performed non-LTE calculations for O~I with the classical plane-parallel (1D) model atmospheres for a set of stellar parameters corresponding to stars of spectral types from A to K. The multilevel model atom produced by Przybilla et al. (2000) was updated using the best theoretical and experimental atomic data available so far. Non-LTE leads to strengthening the O~I lines, and the difference between the non-LTE and LTE abundances (non-LTE correction) is negative. The departures from LTE grow toward higher effective temperature and lower surface gravity. In the entire temperature range and log g = 4, the non-LTE correction does not exceed 0.05 dex in absolute value for lines of O~I in the visible spectral range. The non-LTE corrections are significantly larger for the infrared\\ O~I 7771-5~\AA\ lines and reach -1.9 dex in the model atmosphere with T$_{\rm eff}$ = 10000 K and log~g~=~2. To differentiate the effects of inelastic collisions with electrons and neutral hydrogen atoms on the statistical equilibrium (SE) of O~I, we derived the oxygen abundance for the three well studied A-type stars Vega, Sirius, and HD 32115. For each star, non-LTE leads to smaller difference between the infrared and visible lines. For example, for Vega, this difference reduces from 1.17 dex in LTE down to 0.14 dex when ignoring LTE. To remove the difference between the infrared and visible lines completely, one needs to reduce the used electron-impact excitation rates by Barklem (2007) by a factor of 4. A common value for the scaling factor was obtained for each A-type star. In the case of Procyon and the Sun, inelastic collisions with H~I affect the SE of O~I, and agreement between the abundances from different lines is achieved when using the Drawin's formalism to compute collisional rates. The solar mean oxygen abundance from the O~I 6300, 6158, 7771-5, and 8446 \AA\ lines is $\log \varepsilon = 8.74\pm0.05$, when using the MAFAGS-OS solar model atmosphere and $\log \varepsilon = 8.78\pm0.03$, when applying the 3D corrections taken from the literature. 

 \end{abstract}
\maketitle

\section{Introduction}

    Oxygen abundances of cool stars with different metallicities are important for understanding the galactic chemical evolution. The O~I 7771-4 \AA\ line is observed
for stars in a wide range of spectral types from B
to K, and this is the only set of atomic oxygen lines that is
observed in the spectra of metal-poor stars.
Previously, many authors have shown that different O~I
lines give different abundances, with the infrared (IR)
O~I 7771-5 \AA\ triplet exhibiting a systematically higher
abundance, occasionally by an order of magnitude
higher than the remaining lines do. The reason is that
the IR O~I lines are formed under conditions far from
local thermodynamic equilibrium (LTE). The oxygen
abundance was first determined by abandoning LTE
(within the so-called non-LTE approach) by Kodaira
and Tanaka (1972) and Johnson (1974) for stars and
by Shchukina (1987) for the Sun. Subsequently,
more complex model atoms for O~I were constructed
by Kiselman (1991), Carlsson and Judge (1993),
Takeda (1992), Paunzen et al. (1999), Reetz (1999),
Mishenina et al. (2000), and Przybilla et al. (2000).
Allowance for non-LTE effects leads to a strengthening of lines and, consequently, to a decrease in
the abundance derived from these lines. For A-type
main-sequence stars and F supergiants, even if the
departures from LTE are taken into account, the IR
O~I lines still give a systematically higher abundance
(by 0.25 dex for Vega; Przybilla et al. 2000) than do
the visible lines, for which the departures from LTE
are small (< 0.05 dex in absolute value).

  Solar abundance of oxygen is a key parameter for the studies of solar physics. Having
considered atomic and molecular lines in the solar
spectrum, Asplund et al. (2004) achieved agreement between the abundances from different lines
using a three-dimensional (3D) model atmosphere
based on hydrodynamic calculations. In Asplund
et al. (2004), the mean abundance from atomic
and molecular lines is $\eps$ = 8.66 $\pm$ 0.05;                                                     
subsequently, these authors obtained $\eps$ = 8.69 by the
same method (Asplund et al. 2009). Here, $\eps$ = log$(n_{elem}/n_{\rm{H}})$+12.
 This value turned out to be
lower than $\eps$ = 8.93 $\pm$ 0.04 obtained previously by
Anders and Grevesse (1989) from OH molecular lines
using the semi-empirical model atmosphere
 of Holweger and Mueller 1974 (HM74). It should be noted
that solar models constructed
with the chemical composition from Anders and
Grevesse (1989) described well the sound speed
and density profiles inferred from helioseismological
observations. A downward revision of the oxygen abundance by
0.27 dex led to a discrepancy between the theory and
observations up to 15 $\sigma$ (Bahcall and Serenelli 2005).
Delahaye and Pinsonneault (2006) showed that
the depth of the convection zone and the helium
abundance deduced from observations allow a lower
limit to be set on the surface oxygen abundance,
$\eps$ = 8.86 $\pm$ 0.05. This value is larger than that
from Asplund et al. (2009) by 0.17 dex (more than
3$\sigma$) and that from Caffau et al. (2008) by 0.10 dex
(2$\sigma$). The authors suggested that such discrepancies
could be due to imperfections of modeling the stellar
atmospheres and line formation. In their next paper,
Pinsonneault and Delahaye (2009) analyzed the errors in analyzing the solar oxygen lines and concluded
that the atmospheric abundance error could reach
0.08 dex. This is greater than that given by Asplund
et al. (2009, 0.05 dex) and is close to the error in
Caffau et al. (2008, 0.07 dex).

    Why do the abundances derived by Asplund
et al. (2004, 2009) and Anders and Grevesse (1989)
differ? The molecular lines are known to be very
sensitive to the temperature distribution in a model
atmosphere. In the surface layers where the lines
originate, the temperature in 3D model atmospheres
is lower than that in classical 1D models, MARCS
(Gustafsson et al. 2008) and HM74. Therefore,
the molecular lines computed with a 3D model are
stronger, while the abundances inferred from them
are lower.

    For the Sun and cool stars, the non-LTE abundance
derived from atomic lines can be inaccurate due to
the uncertainty in calculating poorly known collisions with H ~I atoms. Accurate
quantum-mechanical calculations of collisions with
hydrogen atoms are available only for a few atoms:
for Li I (Belyaev and Barklem 2003), Na I (Belyaev
et al. 1999, 2010; Barklem et al. 2010), and Mg I
(Barklem et al. 2012). Since there are no
accurate calculations and laboratory measurements of the
corresponding cross sections, these are calculated
from the formula derived by Steenbock and Holweger
(1984) using the formalism of Drawin (1968, 1969).
The authors themselves estimate the accuracy of the
formula to be one order of magnitude. A scaling factor
(\kH) that can be found by reconciling the abundances
from lines with strong and weak departures from LTE
is usually introduced in this formula. For example,
for Na I (Allende Prieto et al. 2004), Fe I-Fe II
(Mashonkina et al. 2011), and Ca I-Ca II (Mashonkina
et al. 2007) atoms, these scaling factors are 0,
0.1, and 0.1, respectively. For oxygen, Allende Prieto
et al. (2004) and Pereira et al. (2009) obtained \kH~=~1
by investigating the changes of the O~I line profiles in
different regions of the solar disk; Takeda (1995) obtained
the same result from O~I lines in the solar flux spectrum. For the Sun and cool stars, agreement
between the abundances determined from different O~I
lines can be achieved by choosing a scaling factor.
Caffau et al. (2008) performed non-LTE calculations
for O~I with \kH~=~0, 1/3, and 1. The higher the
efficiency of the collisions with hydrogen atoms, the
smaller the departures from LTE and the higher the
abundance. For example, for the IR O~I 7771 \AA\ line,
the abundance inferred with \kH~=~1 is higher than
that with \kH~=~0 by 0.12 dex. Asplund et al. (2009)
disregarded the hydrogen collisions (\kH~=~0). Note
that when using the same methods (non-LTE, \kH~=~0, 3D model atmospheres) and common atomic lines, Caffau et al. (2008) and
Asplund et al. (2004) give $\eps$~=~8.73 and $\eps$~=~8.64, respectively. At such a difference between different
authors (0.09 dex), the solar oxygen abundance
cannot be considered a well-determined quantity.
The accuracy of non-LTE calculations depends
not only on the reliability of the data for collisions with
hydrogen atoms but also on the data for collisions with electrons.
To improve the accuracy of non-LTE results,
Barklem (2007) calculated new cross sections for the
excitation of O~I transitions under collisions with electrons.
These data were used by Fabbian et al. (2009)
to determine the abundance and to analyze the departures
from LTE for IR lines in cool metal-poor stars
with an effective temperature T$_{\rm eff}$ = 4500 -- 6500 K. It
is well known (see, e.g., Sneden et al. 1979) that
in LTE the oxygen overabundance with respect to iron
increases with decreasing metallicity, so that [O/Fe]
reaches 0.5 dex already at metallicity [Fe/H] = -- 1.
(Here, the universally accepted designation [X/Y]= log($n_{X}/n_{Y})_{*}$\;--\;log($n_{X}/n_{Y})_{\odot}$ is used.) The problem of
oxygen overabundances was considered using a non-LTE approach by Abia and Rebolo (1989), Mishenina
et al. (2000), and Nissen et al. (2002). Allowance
for the departures from LTE does not eliminate the
[O/Fe] overabundance, but it is reduced. Nissen
et al. (2002) disregarded the hydrogen collisions and
obtained a linear growth of [O/Fe] from 0 to 0.3 at
0 > [Fe/H] > -- 1, which was replaced by an almost
constant ratio at -- 1 > [Fe/H] > -- 2.7. The same
result was obtained by Fabbian et al. (2009) with the
data from Barklem (2007), but with allowance made
for the hydrogen collisions with \kH~=~1. In this case,
the introduction of hydrogen collisions compensates
for the decrease in the rate coefficient for collisions
with electrons in Barklem (2007) in comparison with
the previous data. With Barklem's new data and
without any hydrogen collisions, the non-LTE effects
increase. Thus, $\Delta_{non-LTE}$ = -- 1.2 dex at [Fe/H] =
-- 3.5 (T$_{\rm eff}$ = 6500 K, log g = 4) and the [O/Fe] ratio
decreases with decreasing metallicity (Fabbian
et al. 2009), which cannot be explained by the existing
nucleosynthesis models.

The goal of our paper is to obtain a reliable method
of determining the oxygen abundance for stars of
spectral types from A to K from different O~I lines.
First, we checked how using the new data from
Barklem (2007) affects the oxygen abundance
determination for hot stars with T$_{\rm eff}$ > 7000 K, for
which the statistical equilibrium of O~I does not
depend on collisions with neutral hydrogen atoms. We
analyzed the spectra of four stars with reliably determined
parameters. Then we 
estimated \kH\ by analyzing the O~I lines for the Sun and
Procyon. We performed an independent analysis of
the solar O~I lines using the most accurate available
atomic data and modeling methods to understand the
possible causes of the uncertainties in determining the solar
oxygen abundance and to try to refine its value.  We describe the model atom and
the mechanism of departures from LTE in Section 2,
the methods and codes used in Section 3, and the
observations and stellar parameters in Section 4.
We present the results for hot stars in Section 5
and for the Sun and Procyon in Section 6. We
also calculated the non-LTE corrections for a grid
of model atmospheres; these are given in Section 7.

\section{The model atom and statistical equilibrium of oxygen}

\begin{figure}  
\resizebox{100mm}{!}{\includegraphics{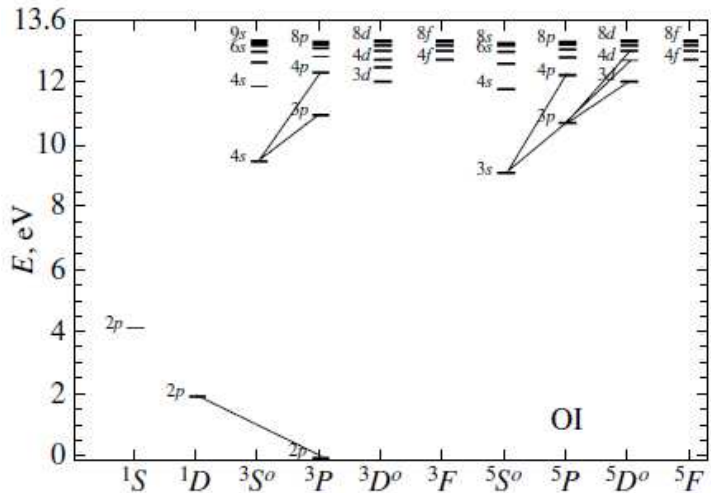}}
\caption{O~I model atom. The straight lines correspond to the transitions in which the lines under study are formed.} 
\label{test22}
\end{figure}

We used the model atom from Przybilla et al.
(2000). It includes 51 levels of O~I and the O~II ground
state (Fig. 1). The oxygen levels belong to the singlet,
triplet, and quintet terms. The maximum principal
quantum number in this model is n = 10. The model
atom accounts for the radiative and collisional processes
in bound-free and bound-bound transitions.
The atomic data used are described in detail in Przybilla
et al. (2000). Przybilla et al. (2000)
point out that the uncertainty in the data for
collisions with electrons makes the most significant
contribution to the uncertainty in final results of non-LTE calculations
for hot stars. They estimated the accuracy of the
atomic data to be 10\% for radiative transitions and
50\% for collisional ones.

To improve the accuracy of non-LTE calculations,
Barklem (2007) calculated new cross sections for
the excitation of O~I transitions under collisions with
electrons. In the model atom, we made changes for
153 transitions associated with the new theoretical
calculations by Barklem (2007). In the model atom
from Przybilla et al. (2000), the rate coefficients
for these transitions were calculated using either the
formula from Wooley and Allen (1948), or the formula
from van Regemorter (1962), or the data from Bhatia
and Kastner (1995). Table 1 compares the rate coefficients
for several transitions at temperature T =
8750 K and electron density $n_e$=1.3 10$^{15}$ $cm ^{-3}$
for the model atom from Przybilla et al. (2000) (the
P00 column) and our updated model with the data
from Barklem (2007) (the B07 column). The chosen
parameters correspond to the atmosphere of Vega
at depth log($\tau_{5000}$) = -- 1. As an example, we took
several transitions with energies $\Delta E_{ij}$ < 2 eV, because
a change in the rate coefficients with larger and
smaller energy separations affects weakly the statistical
equilibrium of O~I. The rate coefficients increased
by several orders of magnitude for some transitions
and decreased for others. Most importantly, the new
rate coefficients decreased for the transitions with the
largest rate coefficients, which affect most strongly
the statistical equilibrium of O~I. A change in the
data for the transitions from the ground level and the
transitions between triplet and quintet levels does not
affect the result, as was previously shown by Kiselman
(1991) and Carlsson and Judge (1993).
The departures from LTE are commonly characterized
by b-factors, b$_i$ = n$_{i \rm {non-LTE}}$/n$_{i \rm {LTE}}$, where
n$_{i \rm {non-LTE}}$ and n$_{i \rm {LTE}}$ are the populations of the i-th level
in the non-equilibrium and equilibrium cases. Figure 2
shows the behavior of the b-factors for several levels
in the atmospheres of Vega (T$_{\rm eff}$ = 9550 K) and the
Sun (T$_{\rm eff}$ = 5780 K). The qualitative behavior of the
b-factors in the atmosphere
is similar; the only difference is that the departures
from LTE for Vega are stronger due to the higher
T$_{\rm eff}$ and lower log g. Two types of processes compete
between themselves in the stellar atmosphere:
collisional ones, which tend to bring the system to
equilibrium, and radiative ones, which hinder them.
The collisional transition rates depend on local parameters
of the medium, while the radiative ones
depend on the mean intensities of radiation that are non-local quantities in
the upper atmospheric layers, where the photon mean
free path is large. Where the density is high, the
rate of collisional processes is higher than that of
radiative ones; the conditions are close to equilibrium
ones. In deep layers (log($\tau_{5000}$) > 0 for the Sun
and log($\tau_{5000}$) > 1 for Vega), the populations of all
levels do not differ from the equilibrium ones. In
the atmospheres of the Sun and Vega, most of the
O~I atoms are at levels of the main 2p configuration
(the three lower levels in Fig. 1). Even for Vega,
the ratio of the O~I and O~II ground level populations
$\rm{n_{O~I}/n_{O~II}}$ > 10 at depths log($\tau_{5000}$) < 0.3. Therefore,
the populations of all three levels do not differ from the
equilibrium one. The O~I and O~II ground levels are
related by the charge exchange reaction H$^+$~+~O~$\leftrightarrow$~H~+~O$^+$, because the O~I and H~I ionization potentials
are closely spaced (13.62 and 13.60 eV, respectively). As a
result, the O~II level population follows the O~I ground
state population, and the b-factors in the figure merge
together. Spontaneous transitions from the highly
excited levels tightly coupled with the O~II ground
state lead to overpopulation of the lower states. The
 $3s \;^3S^\circ$ and $3s \;^5S^\circ$ levels are overpopulated
more than all other levels.
To understand how the line profile changes under
non-equilibrium conditions, one should take a look
at the behavior of the b-factors for upper (b$_{j}$) and lower
(b$_{i}$) levels at the line formation depth. The line profile
is affected by the deviation of the source function
($S_\nu$) from the Planck function (B$_{\nu}$) and the change in
opacity ($\chi_\nu$) under non-equilibrium conditions. In the
visible range, these quantities depend on the b-factors
as follows:

 $S_\nu \sim B_\nu b_j/b_i$,   \;\;\;\;\;\;\    $\chi_\nu \sim b_i$ 

For the strong IR O~I 7771-5 \AA\ (the $3s \;^5S^\circ$ -- $3p\; ^5P$
transition) lines, whose core is formed at depth
log($\tau_{5000}$) $\simeq$ -- 2,  departures from LTE are large
due to overpopulation of lower-level  and
$b_{3p \;^5P}/b_{3s \;^5S^\circ}$   < 1 (Fig. 2), that leads to the strengthening of the lines.
The visible (3947, 4368, 5330, 6155-9 \AA ) O~I lines
also strengthen. However, since they are weak and
originate in deep layers, non-LTE effects do not lead
to such a dramatic change of the line profiles. The
forbidden 6300 \AA\ line is
immune to departures from LTE.
It is formed in the 2p\;$^3$P - 2p\;$^1$D$^\circ$
transition, whose level populations do not differ from
the equilibrium ones.

\begin{table}[htbp]
\caption{ Rate coefficients for collisional transitions in different
model  atoms for oxygen.}
\label{lines2}
\tabcolsep3.7mm
\begin{center}
\begin{tabular}{ccc}
\hline\noalign{\smallskip}
  Transition  & C$_{lu}$, $c^{-1}cm^{-3}$ (P00)  &  C$_{lu}$, $c^{-1}cm^{-3}$ (B07) \\  
\hline\noalign{\smallskip}
   $2p\;^3P^o$ --$2p\;^1D^o$    & 6.794e5 & 2.357e4 \\
    $3s\;^5S^o$  -- $3s\;^3S^o$   & 5.275e3 & 5.910e6 \\
   $3s\;^5S^o$  -- $3p\;^5P$  & 1.041e7 & 7.611e6 \\
   $3s\;^3S^o$  -- $3p\;^3P$  & 1.573e7 & 8.524e6 \\
   $3p\;^5P$  -- $3p\;^3P$  & 1.255e6 & 1.743e6 \\
 \hline
\end{tabular}
\end{center}
\end{table} %

\begin{figure}  
\resizebox{80mm}{!}{\includegraphics{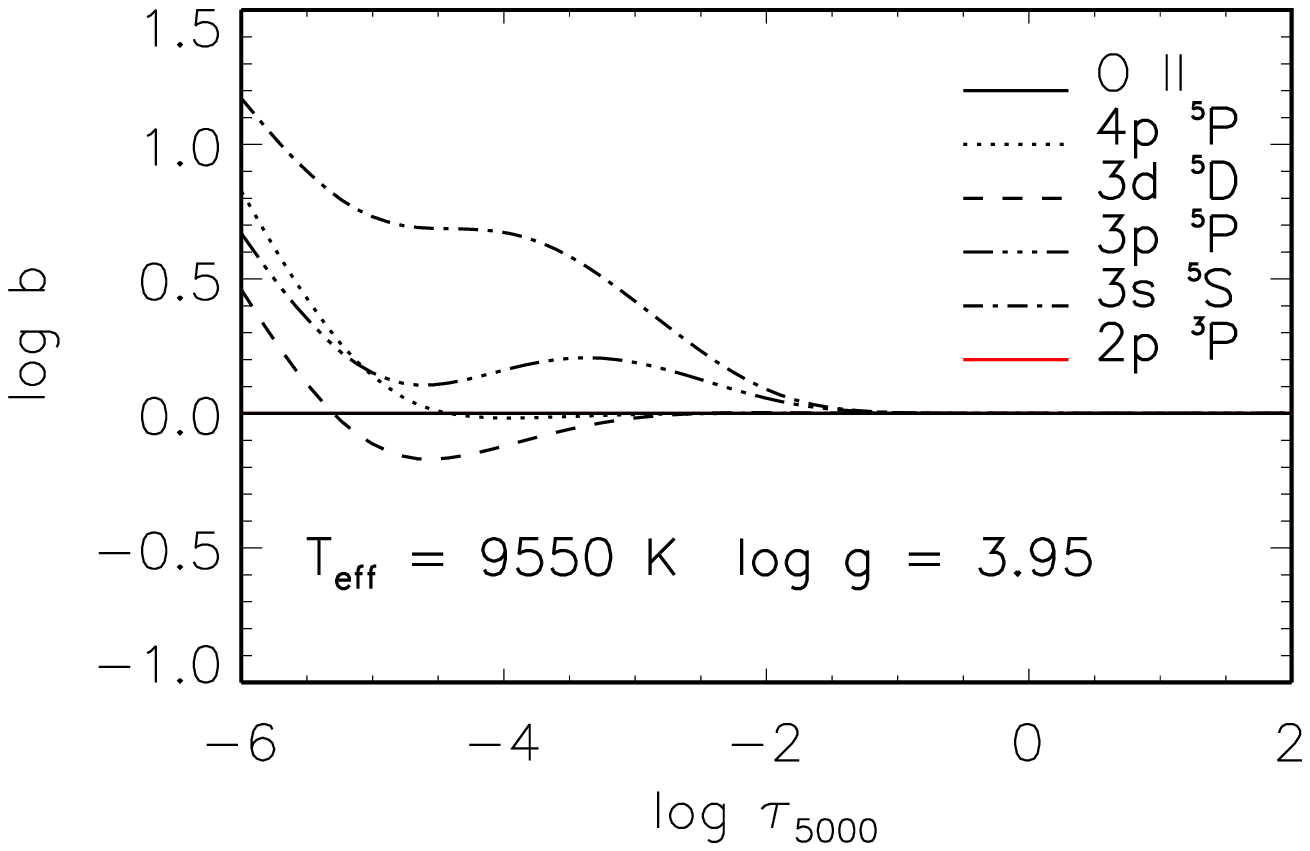}}
\resizebox{80mm}{!}{\includegraphics{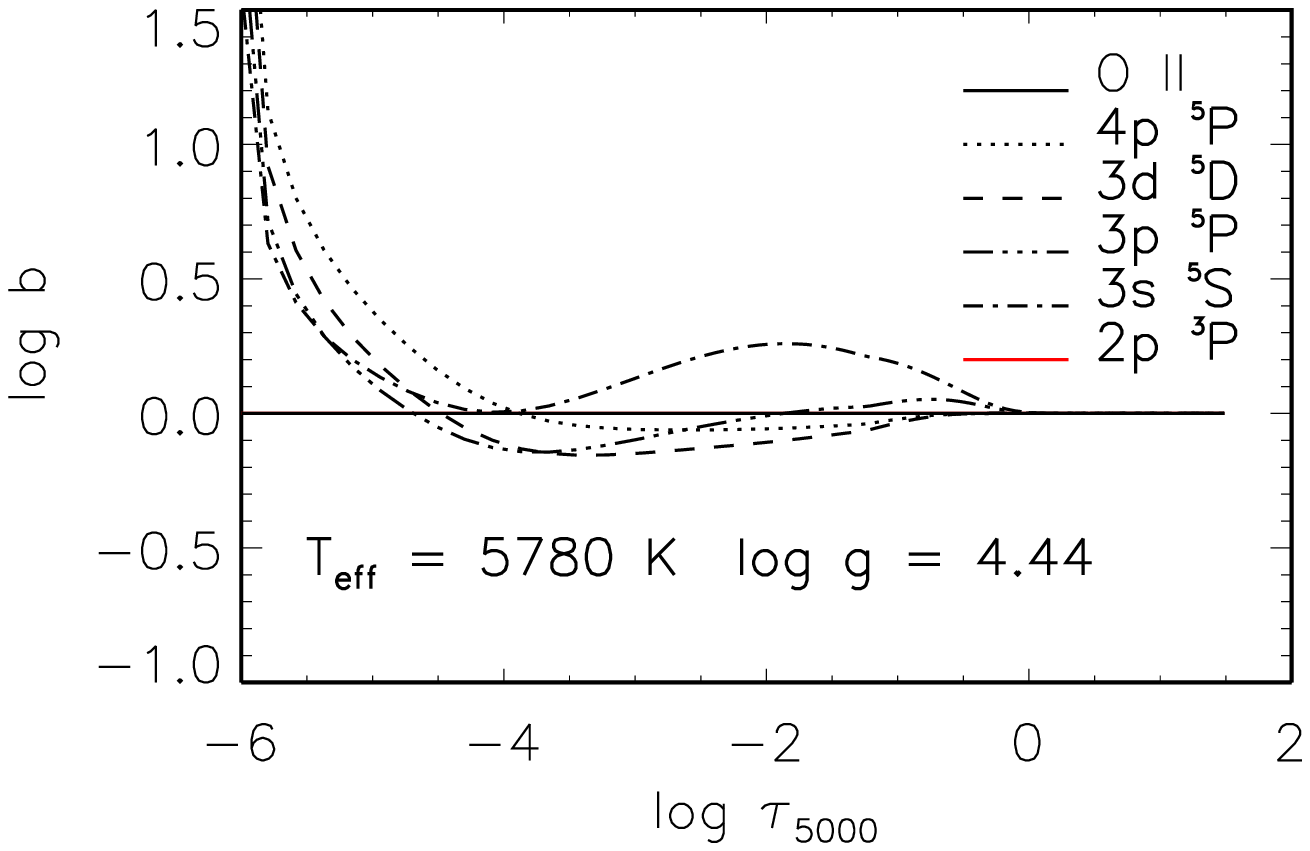}}
\caption{b-factors for oxygen levels in the atmospheres of Vega (left) and the Sun (right).}
\label{vega1}
\end{figure}


\section{The codes and methods}

We calculated the equilibrium and non-equilibrium
level populations using the DETAIL code developed
by Butler and Giddings (1985) and based on the
method of an accelerated $\lambda$ -- iteration. The opacity
in continuum and lines (42 million lines for 90
elements) is taken into account. The output data
from the DETAIL code (level populations) were
then used to compute a synthetic spectrum by different
codes. For stars with T$_{\rm eff}$ < 7500 K, we applied
the SIU (Spectrum Investigation Utility) code
(Reetz 1999), which allows a synthetic spectrum
to be computed in the equilibrium and non-equilibrium
cases. For the remaining stars, we used the
SYNTH3 code (O. Kochukhov), with which the LTE
abundance can be calculated and the line equivalent
widths can be measured. To derive the non-LTE
abundances, we worked with the equivalent widths in
the LINEC code (Sakhibullin 1983). The same code
was used to calculate the non-LTE corrections for a
grid of model atmospheres. A preliminary test shows
that different codes give the same abundance within
0.02 dex.

\section{Observations and stellar parameters}

We used high-spectral-resolution observations
whose characteristics are listed in Table 2. Table 3
gives the stellar parameters taken from the literature.
For Procyon, the effective temperature and surface
gravity were determined, respectively, by the method
of IR fluxes and the astrometric method (Chiavassa
et al. 2012). For Vega and Deneb, the effective
temperatures and surface gravities were determined
from the Balmer line wings and Mg~I and Mg~II
lines (Przybilla et al. 2000; Schiller and Przybilla
2008). For HD 32115, the effective temperature
and surface gravity were determined, respectively, from
the H$_\gamma$ line wings and the Mg~I line wings taking
into account the departures from LTE (Fossati
et al. 2011). For Sirius, we took Kurucz's model
atmosphere (http://kurucz.harvard.edu/stars/ SIRIUS/
ap04t9850g43k0he05y.dat) computed with parameters
close to those derived by Hill and Landstreet
(1993).
We decided to independently determine the microturbulence
for Sirius, because the published values
vary in a wide range between \vt = 1.7 \kms (Hill
and Landstreet 1993) and \vt = 2.2 \kms (Landstreet
2011). The microturbulence is usually determined
by the value at which the dependence of
the elemental abundance derived from individual lines
on equivalent widths vanishes. The most accurate
value of this parameter is obtained when using lines
with a wide range of observed equivalent widths, from
several to 100-150 m\AA. In the spectrum of Sirius,
the lines of ionized iron meet this criterion. To determine
the microturbulence, we selected 27 Fe II
lines with equivalent widths in the range from 9 to
170 m \AA\ and with fairly reliable oscillator strengths.
All lines were carefully checked for possible blending
by comparison with a synthetic spectrum computed
with a sample of lines from VALD (Kupka et al. 1999).
For a more reliable determination of the microturbulence,
we used two sets of oscillator strengths:
theoretical calculations based on the method of orthogonal
operators (Raassen and Uylings 1998) and
a recent compilation by Melendez and Barbuy (2009).
In Fig. 3, the iron abundance from individual lines
is plotted against the equivalent width for \vt = 1.4,
1.7, and 2.0 \kms. For both sets of oscillator
strengths, we obtained a consistent microturbulence,
\vt= 1.7 -- 1.8 \kms, with an error of $\pm$ 0.3 \kms.
Our result is in good agreement with \vt = 1.7 \kms
and \vt = 1.85 \kms from Hill and Landstreet (1993)
and Qiu et al. (2001). The value of \vt = 2 \kms
used in some papers (Lemke 1990; Hui-Bon-Hoa
et al. 1997; Landstreet 2011) probably results from
the fact that small blends were disregarded when
measuring the equivalent widths and determining the
abundance from strong Fe II and Fe I lines.

We used classical 1D model atmospheres computed
with the following codes: LLmodels (Shulyak
et al. 2004) for HD 32115 and Vega, MAFAGSOS
(Grupp et al. 2009) for the Sun and Procyon,
ATLAS12 (R. Kurucz) for Sirius, and ATLAS9 (Kurucz
1994) for Deneb, and to calculate the non-LTE
corrections for a grid of model atmospheres. Using
different codes to compute model atmospheres is admissible,
because our goal is to achieve agreement
between the abundances from different lines for each
star, not to compare the abundances for stars with one
another.

\begin{figure}[htbp]
\hspace{-6mm}
\resizebox{120mm}{!}{\rotatebox{270}{\includegraphics{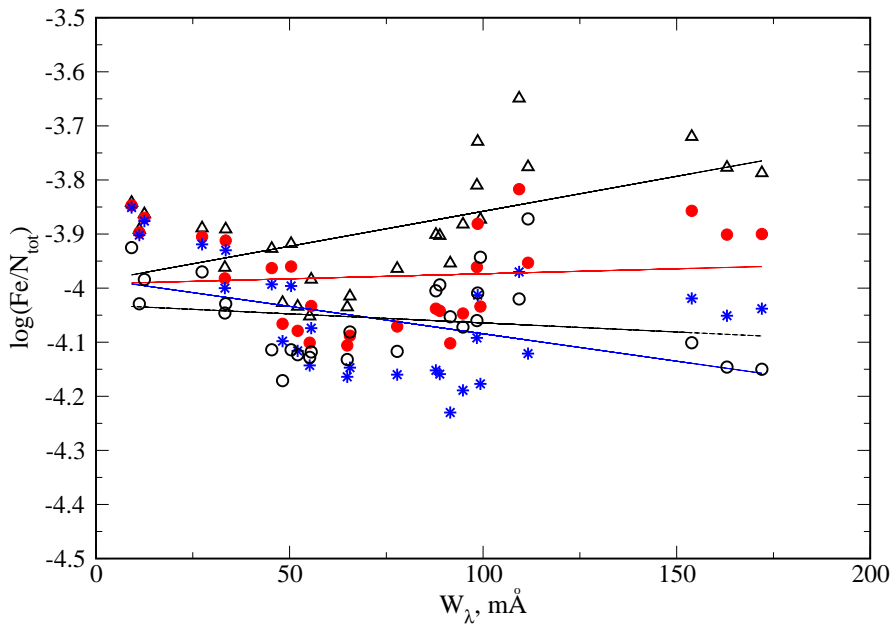}}}
\caption{Iron abundance from individual lines versus equivalent width. The data obtained using the oscillator strengths from
Raassen and Uylings (1998) for \vt = 1.4, 1.7 and 2.0\,\kms, are indicated by the open triangles, filled circles, and asterisks,
respectively. The open circles correspond to the data obtained with \vt = 1.7\,\kms\ and the oscillator strengths from Melendez
and Barbuy (2009). The straight lines indicate a linear fit. } 
\label{sirius_vt}
\end{figure}

\begin{table}[htbp]
\caption{ Characteristics of the observations.}
\label{lines}
\tabcolsep3.7mm
\begin{center}
\begin{tabular}{llccl}
\hline\noalign{\smallskip}
 Star & HD&$\lambda/\Delta\lambda$& $S/N >$& Source\\
\hline\noalign{\smallskip}
  Sun   &             &300000&300& Kurucz et al. (1984)\\
  Procyon & 61421   &65000&200&  Korn et al. (2003) \\
                 & 32115   &30000& 150& Bikmaev et al. (2002)\\
  Vega        & 172167 &60000&200& A. Korn, private communication\\
  Sirius   & 48915   &70000&500& Furenlid et al. (1995)\\
  Deneb     & 197345 &42000&200&ELODIE http://atlas.obs-hp.fr/elodie/\\ 
\hline
\end{tabular}
\end{center}
\end{table}

\begin{table}[htbp]
\caption{ Stellar parameters.}
\label{lines1}
\tabcolsep3.0mm
\begin{center}
\begin{tabular}{llccrcccl}
\hline\noalign{\smallskip}
 Star & HD& T$_{\rm eff}$ & $\log g$ & [Fe/H] & \vt &  $\eps$ & $\sigma$& Source \\ 
\hline\noalign{\smallskip}
Sun   &             &  5777 & 4.44 &   0.0 & 0.9 & 8.74 & 0.03 & \\
 Procyon & 61421   &  6590 & 4.00 &   0.0 & 1.8 &   8.73 & 0.03&  Chiavassa et al. (2012)\\
                 & 32115   &  7250 & 4.20 &   0.0 & 2.3 & 8.78 & 0.09&  Fossati et al. (2011)\\
 Vega        & 172167 & 9550 & 3.95 & --0.5 & 2.0 & 8.59&0.01&  Przybilla et al. (2000)\\ 
 Sirius   & 48915   & 9850 & 4.30 &   0.4 & 1.8$^*$ & 8.42 & 0.03& Hill and Landstreet (1993)\\
 Deneb     & 197345 & 8525 & 1.10 & --0.2 &8.0 &  8.58  &  0.01& Schiller and Przybilla (2008)\\ 
\hline
\end{tabular}
\end{center}
$^*$ Determined in this paper
\end{table}

\section{The oxygen abundance in hot stars}

The list of lines from which the oxygen abundance
in the stars was determined is given in Table 4. The
atomic data for the transitions, namely, wavelengths ($\lambda$), oscillator
strengths (log gf), lower-level excitation energies
(E$_{exc}$), radiative damping constants (log $\gamma_{rad}$),
and quadratic Stark (log $\gamma_4$) and van der Waals
(log $\gamma_6$) broadening constants per particle at T =
10$^4$~K were taken from VALD (Kupka et al. 1999),
except for the oscillator strength for the 6300~\AA\ line
calculated by Froese Fisher et al. (1998).
For Vega, we performed our non-LTE calculations
with the model atom from Przybilla et al. 2000 (P00) and updated one with data from
Barklem 2007. The derived
abundances are given in Table 5. We separated
the lines into two groups: with small ($\Delta_{non-LTE} \simeq $ --
0.05 dex) and large ($\Delta_{non-LTE} \simeq $ --1 dex) departures
from LTE. Here, the non-LTE correction $\Delta_{non-LTE} =
\eps_{non-LTE} - \eps_{LTE}$. In LTE, the abundance difference
between the two groups is 1.23 dex. In our
non-LTE calculations with P00 atom model, the difference
decreases considerably and is 0.33 dex. With the
new data, the situation changes for the better, but a
difference of 0.14 dex still remains. The departures
from LTE have increased, because decreasing of the rate coefficients
for the most important transitions for statistical
equilibrium (with $C_{lu} \sim 10^7$ $c^{-1}cm^{-3}$ or more). For
example, for the $3s\; ^5S^\circ - 3p\; ^5P$ transition, in which
the 7771-5 \AA\ lines are formed, the rate coefficient
decreased by a factor of 1.4. Test calculations showed
that such a change in the rate coefficient only for
one transition makes the largest contribution to the
change in the non-LTE correction for the 7771-5 \AA\
lines. 
Przybilla et al.(2000) et al. (2000) estimated the
accuracy of the atomic data to be 10 \% for radiative
transitions and 50 \% for collisional ones. Therefore,
we believe that the uncertainty of our non-LTE calculations
for Vega is probably related only to the
data for collisions with electrons. We decided to introduce a scaling factor for the rate coefficients from
Barklem (2007), derived from the requirement, that different lines must give the same abundance.
The obtained scaling factor \se~=~0.25, the abundance difference 0.02 dex
between the two groups of lines for Vega is achieved.

Thereafter, similar non-LTE calculations were
performed for HD 32115 and Sirius; the results are
presented in Tables \ref{lines4} and \ref{lines5}. For HD 32115,
the group of lines with large non-LTE corrections
includes not only the 7771-5 \AA\ lines but also the 8446
and 9266 \AA\ lines. With the data from Barklem (2007),
the difference between the abundances from different
groups of lines is 0.09 and 0.14 dex for HD 32115
and Sirius, respectively, and reduces down to 0.04 and 0.02 dex when using a scaling factor \se~=~0.25.

\begin{table}[htbp]
\caption{ List of lines with atomic parameters.}
\label{lines3}
\tabcolsep3.7mm
\begin{center}
\begin{tabular}{llccccc}
\hline\noalign{\smallskip}
$\lambda$, \AA   & E$_{exc}$, н‚   & log gf & log $\gamma_{rad}$&log $\gamma_4$&  log  $\gamma_6 $&   Transition   \\    
\hline\noalign{\smallskip}
3947.29279   &  9.146 & --2.096 &6.660& --4.700& --7.957&   $3s\; ^5S^\circ$  -- $4p\; ^5P$ \\
3947.48274   &  9.146 & --2.244 &6.660& --4.700& --7.957&   $3s\; ^5S^\circ$  -- $4p\; ^5P$ \\
3947.58272   &  9.146 & --2.467 &6.660& --4.700& --7.957&   $3s\; ^5S^\circ$  -- $4p\; ^5P$ \\
 4368.19244   &  9.521 & --2.665 & 8.760& --4.680&  --7.946& $3s\; ^3S^\circ$  -- $4p\; ^3P$ \\
 4368.24242   &  9.521 & --1.964 & 8.760& --4.680&  --7.946& $3s\; ^3S^\circ$  -- $4p\; ^3P$ \\
 4368.26242   &  9.521 & --2.186 & 8.760& --4.680&  --7.946& $3s\; ^3S^\circ$  -- $4p\; ^3P$ \\
 5330.72716   & 10.741 & --2.415 &7.550& --3.430&   -- & $3p\; ^5P$  -- $5d\; ^5D^\circ$ \\
5330.73716   & 10.741 & --1.570 &7.550& --3.430&  --& $3p\; ^5P$  -- $5d\; ^5D^\circ$ \\
5330.73716   & 10.741 & --0.984 &7.550& --3.430&   --& $3p\; ^5P$  -- $5d\; ^5D^\circ$ \\
 6155.96637   & 10.741 & --1.363 &7.600& --3.960  &--6.860 &  $3p\; ^5P$ -- $4d\; ^5D^\circ$ \\
 6155.96637   & 10.741 & --1.011 &7.610& --3.960  &--6.860 &  $3p\; ^5P$ -- $4d\; ^5D^\circ$ \\
 6155.98637   & 10.741 & --1.120 &7.610& --3.960  &--6.860 &  $3p\; ^5P$ -- $4d\; ^5D^\circ$ \\
 6156.73616   & 10.741 &--1.488 &7.610& --3.960  &--6.860 &  $3p \;^5P$  -- $4d\; ^5D^\circ$  \\
 6156.75616   & 10.741 & --0.899 &7.610& --3.960  &--6.860 &  $3p \;^5P$  -- $4d\; ^5D^\circ$  \\
 6156.77615   & 10.741 & --0.694 &7.620& --3.960  &--6.860 &  $3p \;^5P$  -- $4d\; ^5D^\circ$  \\
 6158.14579   & 10.741 & --1.841 &7.620& --3.960  &--6.860 &  $3p\; ^5P$  -- $4d\; ^5D^\circ$  \\
 6158.17578   & 10.741 & --0.996 &7.620& --3.960  &--6.860 &  $3p\; ^5P$  -- $4d\; ^5D^\circ$  \\
 6158.18577   & 10.741 & --0.409 &7.610& --3.960  &--6.860 &  $3p\; ^5P$  -- $4d\; ^5D^\circ$  \\
 6300.30400   &  0.000 & --9.720$^*$ &--2.170&  --  & -- &  $2p\; ^3P$ -- $2p\; ^1D$\\
 7771.94130   &  9.146 &  0.369 &7.520& --5.550 & --7.469 &  $3s\; ^5S^\circ$  -- $3p\; ^5P$\\
 7774.16071   &  9.146 &  0.223 &7.520& --5.550 & --7.469 &  $3s\; ^5S^\circ$  -- $3p\; ^5P$ \\
 7775.39037   &  9.146 &  0.001 &7.520& --5.550 & --7.469 & $3s\; ^5S^\circ$  -- $3p\; ^5P$\\
 8446.24912   &  9.521 &  --0.463 &8.770& --5.440&   --& $3s\; ^3S^\circ$  -- $3p\; ^3P$\\
8446.35909   &  9.521 &  0.236 &8.770& --5.440&   --& $3s\; ^3S^\circ$  -- $3p\; ^3P$\\
8446.75898   &  9.521 &  0.014 &8.770& --5.440&   --& $3s\; ^3S^\circ$  -- $3p\; ^3P$\\
 9265.82735   & 10.741 &  --0.719 &7.900& --4.950 & -- &  $3p\; ^5P$  --$3d\; ^5D^\circ$\\
9265.92732   & 10.741 &  0.126 &7.940& --4.950 &  -- &  $3p\; ^5P$  --$3d\; ^5D^\circ$\\
9266.00730   & 10.741 &  0.712  &7.880& --4.950 & --&  $3p\; ^5P$  --$3d\; ^5D^\circ$\\
\hline
\end{tabular}
\end{center}
$*$ -- The data from NIST
\end{table}  %

\begin{table}[htbp]
\caption{Oxygen abundances for Vega and HD 32115.}
\label{lines4}
\tabcolsep2.0mm
\begin{center}
\begin{tabular}{cccccccc}
\hline\noalign{\smallskip}
$\lambda$ & $\eps_{LTE}$ & $\eps_{non-LTE}$  & $\Delta_{non-LTE}$  & $\eps_{non-LTE}$  & $\Delta_{non-LTE}$ & $\eps_{non-LTE}$  & $\Delta_{non-LTE}$  \\
\AA &  & P00 &  P00 &  B07 &  B07 &  1/4(B07)  &   1/4(B07)   \\
\hline
Vega&&&&&&&\\
\hline
 5330 &  8.61 &  8.60 & --0.01 &  8.59 & --0.02 &  8.59 & --0.02 \\
 6155 &  8.64 &  8.62 & --0.02 &  8.61 & --0.03 &  8.61 & --0.03 \\
 6156 &  8.64 &  8.62 & --0.02 &  8.60 & --0.04 &  8.60 & --0.04 \\
 6158 &  8.64 &  8.62 & --0.02 &  8.60 & --0.04 &  8.60 & --0.04 \\
 Mean& 8.63 &  8.62 &        &  8.60 &        &  8.60 &        \\
 7771 &  9.84 &  8.96 & --0.88 &  8.74 & --1.09 &  8.58 & --1.26 \\
 7774 &  9.83 &  8.94 & --0.89 &  8.73 & --1.10 &  8.57 & --1.26 \\
 7775 &  9.80 &  8.95 & --0.85 &  8.74 & --1.05 &  8.59 & --1.21 \\
 Mean& 9.86 &  8.95 &        &  8.74 &        &  8.58 &        \\
\hline
HD 32115&&&&&&&\\
\hline
3947 &  8.77 &  8.76 & --0.01 &  8.77 &  0.00 &  8.77 &  0.00 \\
 4368 &  8.77 &  8.77 & --0.01 &  8.77 &  0.00 &  8.77 &  0.00 \\
 6155 &  8.76 &  8.74 & --0.02 &  8.73 & --0.03 &  8.73 & --0.03 \\
 6158 &  8.83 &  8.80 & --0.03 &  8.79 & --0.04 &  8.79 & --0.04 \\
 Mean& 8.78 &  8.77 &        &  8.76 &        &  8.76 &        \\
 7771 &  9.60 &  9.10 & --0.50 &  8.96 & --0.64 &  8.92 & --0.68 \\
 7774 &  9.59 &  9.09 & --0.50 &  8.93 & --0.66 &  8.89 & --0.70 \\
 7775 &  9.47 &  8.99 & --0.48 &  8.94 & --0.63 &  8.81 & --0.66 \\
 8446 &  9.34 &  8.86 & --0.48 &  8.73 & --0.61 &  8.64 & --0.70 \\
 9266 &  9.02 &  8.79 & --0.23 &  8.71 & --0.31 &  8.71 & --0.30 \\
 Mean& 9.40 &  8.97 &        &  8.85 &        &  8.80 & \\
\hline
\end{tabular}
\end{center}
\end{table}  %

\begin{table}[htbp]
\caption{ Oxygen abundance for Sirius.}
\label{lines5}
\tabcolsep3.7mm
\begin{center}
\begin{tabular}{rlccrc}
\hline\noalign{\smallskip}
$\lambda$  & $\eps_{LTE}$ & $\eps_{non-LTE}$  & $\Delta_{non-LTE}$ & $\eps_{non-LTE}$  & $\Delta_{non-LTE}$ \\
\AA  &  &  B07 &   B07 &  1/4(B07)  & 1/4(B07)   \\

\hline\noalign{\smallskip}
5330 &  8.49& 8.47 & --0.02 &  8.47 & --0.02 \\
 6155 & 8.44&  8.42 & --0.02 &  8.41 & --0.03 \\
 6156 & 8.44&  8.41 & --0.03 &  8.41 & --0.03 \\
 6158 & 8.44 & 8.42 & --0.02 &  8.42 & --0.02 \\
 Mean & 8.45 &   8.43 &        &  8.43 &        \\
 7771  & 9.46 &   8.62 & --0.82 &  8.45 & --1.01 \\
 7774  & 9.38 &   8.57 & --0.78 &  8.41 & --0.96 \\
 7775 &  9.21  &  8.53 & --0.66 &  8.38 & --0.83 \\
 Mean& 9.35 &  8.57 &        &  8.41 &        \\
\hline
\end{tabular}
\end{center}
\end{table}  %

\begin{figure}  
\resizebox{80mm}{!}{\includegraphics{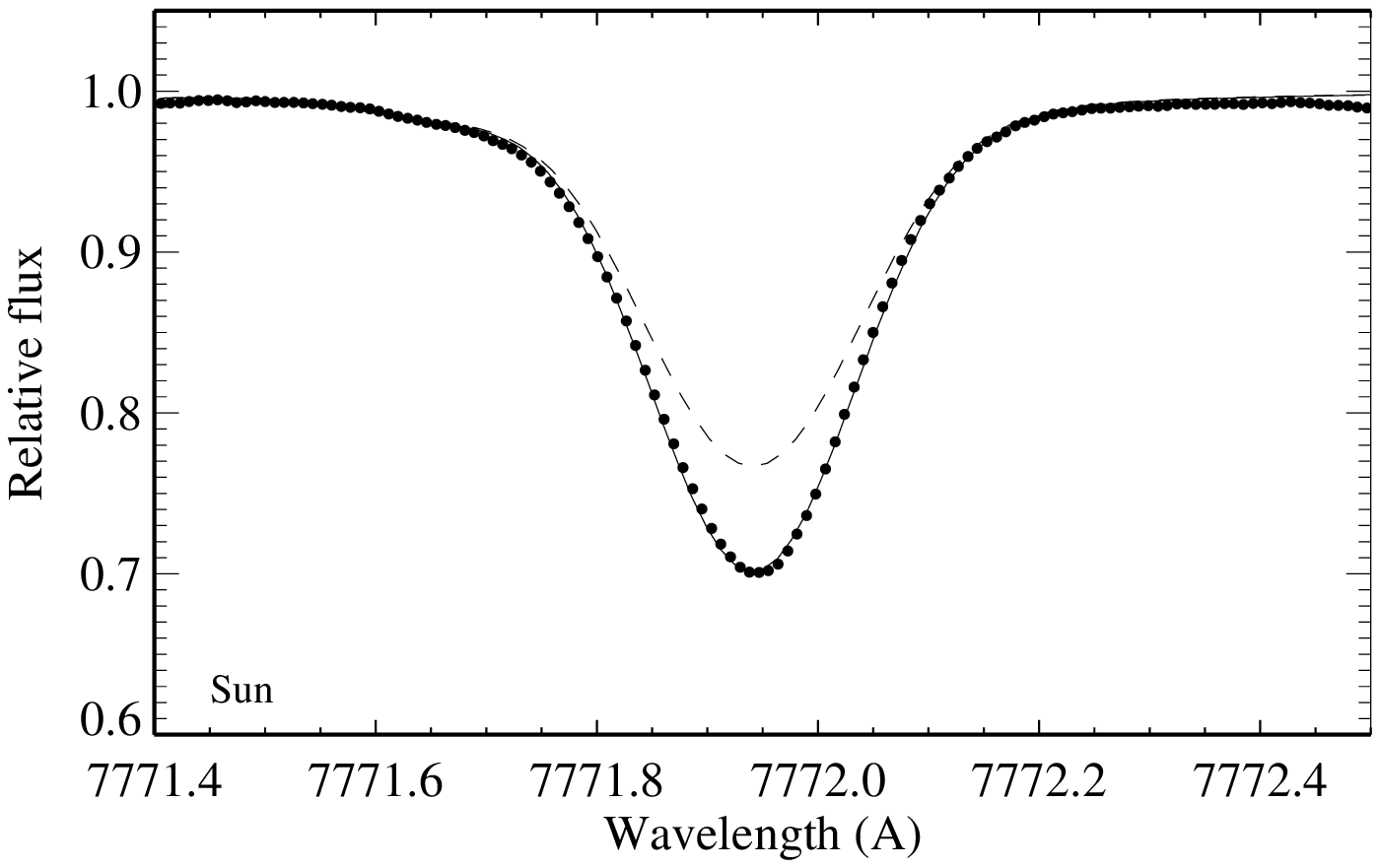}}
\resizebox{80mm}{!}{\includegraphics{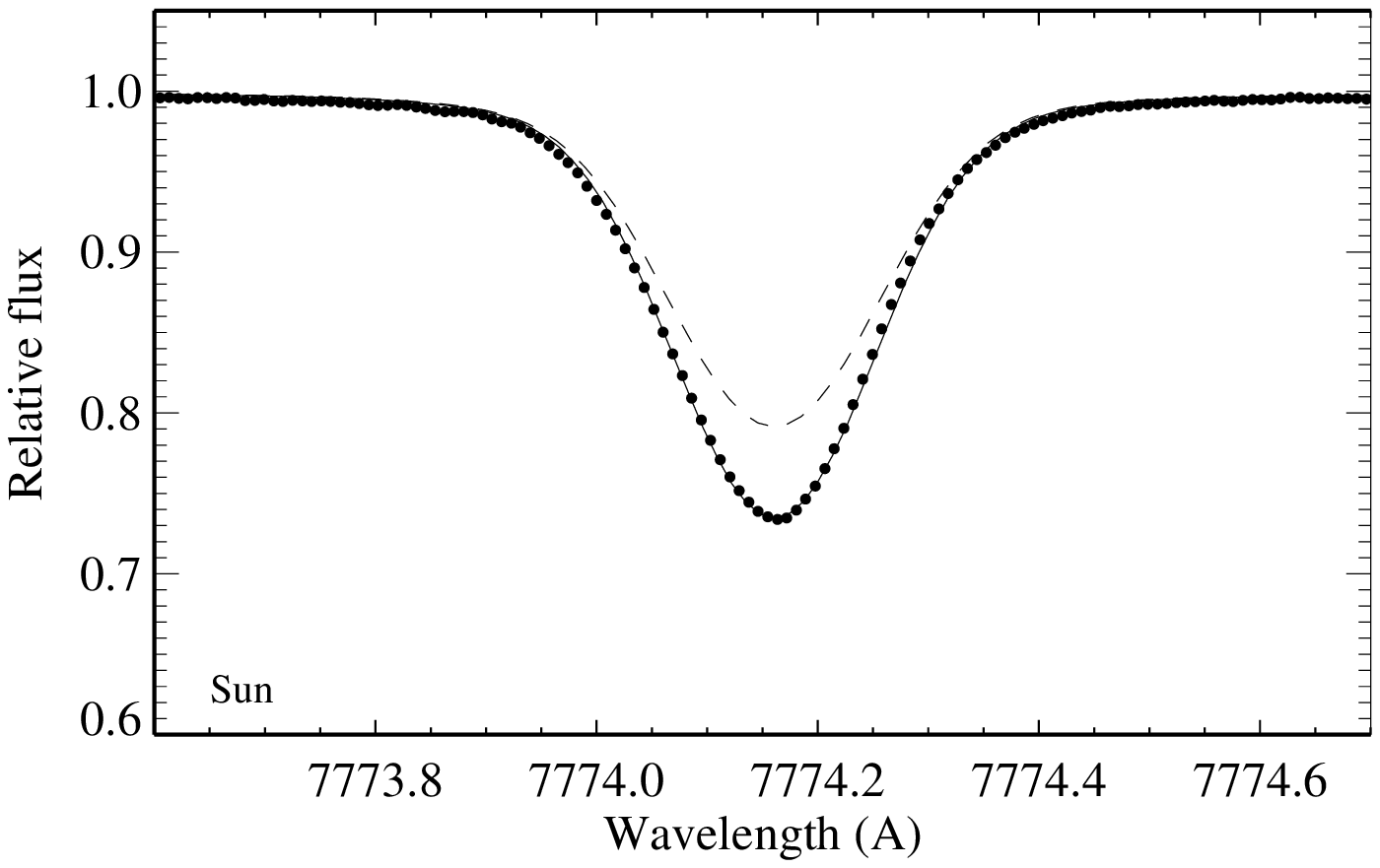}}
\resizebox{80mm}{!}{\includegraphics{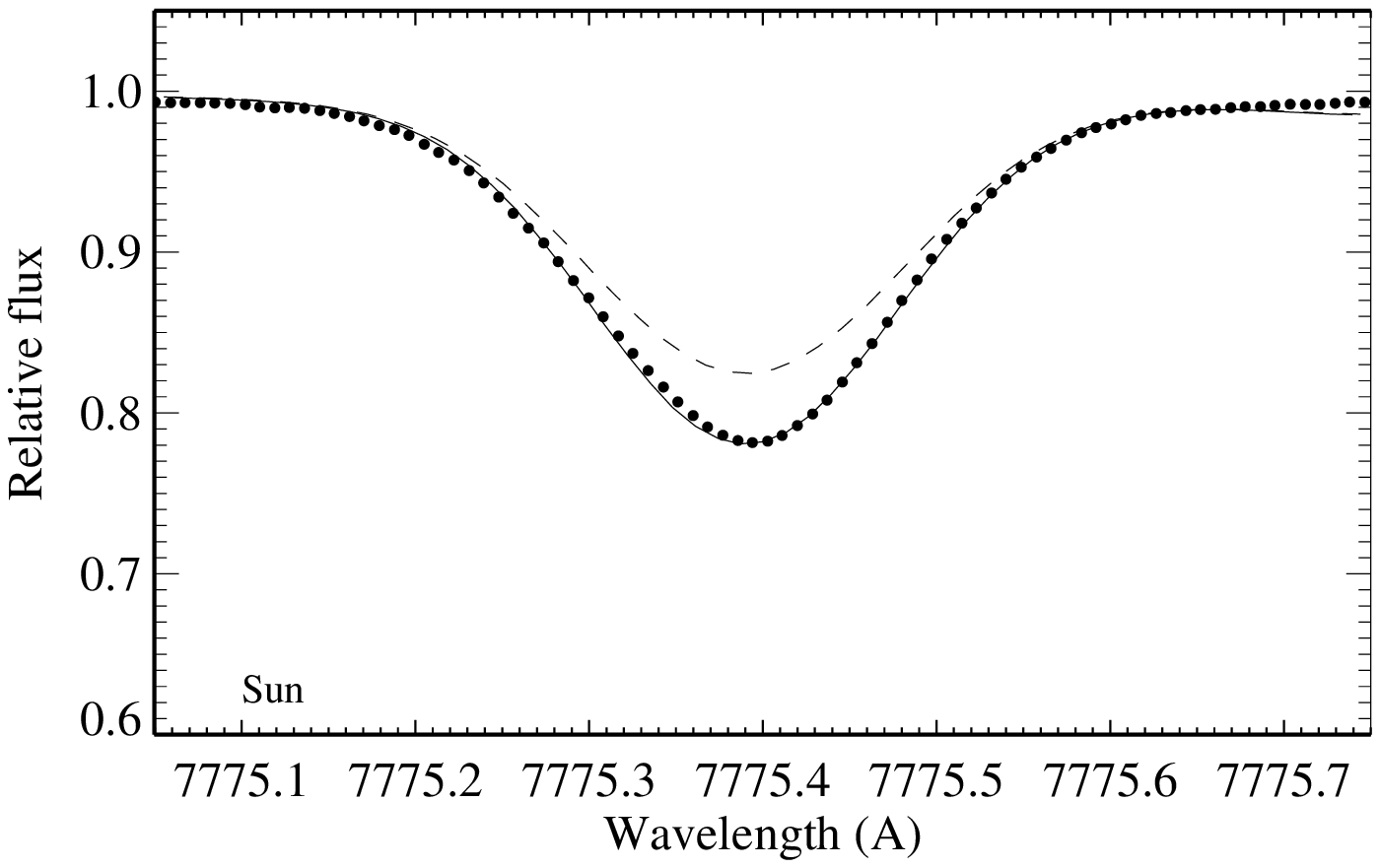}}
\caption{The O~I IR lines in the solar spectrum. The circles, the dotted line, and the solid line indicate the observed spectrum,
the synthetic lte-spectrum, and the non-LTE synthetic spectrum with the oxygen abundance $\eps$ = 8.75 respectively.}
\label{777}
\end{figure}

\section{The solar oxygen abundance}

The abundance
was derived from visible (6300, 6158 \AA ) and
IR (7771-5, 8446 \AA)  (see Fig. \ref{777}) O~I lines (see Table 7). There
is no agreement between the visible lines: in LTE the
difference between the abundances obtained from the
6300 and 6158 \AA\ lines is very large, 0.17 dex. In
the non-LTE case, the difference reduces only
slightly to 0.15 dex. We believe the abundance from
the 6158~\AA\ line to be reliable, because it is not blended
and is much stronger than the forbidden line. There
are several factors related to the 6300 \AA\ line that could lead
to this discrepancy. Figure 4 shows the profile of this
line in the solar spectrum from the Atlas by Kurucz
et al. (1984). The abundance derived from this line
can not be reffered to as reliable, because O~I 6300.304 is weak and
is blended with the Ni~I 6300.336 \AA\ line. In addition,
there is an uncertainty in the oscillator strength:
VALD and NIST give different values of log~gf~=~--~9.82
and -- 9.72. We use the latter value of log~gf~=~--~9.72
calculated theoretically by Froese Fisher et al. (1998).
It should be noted that with the oscillator strength
from VALD, the abundance difference between the
6300 and 6158 \AA\ lines is much smaller, 0.05 dex.
Here, we use classical 1D model atmospheres, but
the formation of these lines may be affected by 3D-effects. From Caffau et al. (2008)
we took the 3D corrections that the authors recommend
to add to the abundance derived with 1D model
atmospheres. In this case, the difference between the
abundances from visible lines reduces down to 0.05 dex.

In LTE the abundance from IR lines is higher than
that from visible ones by 0.14 dex. In non-LTE, with H~I collisions taken into account, 
this difference $\Delta \eps(IR-vis)$~=~0.04 dex for the model atom by
Przybilla et al. (2000) and completely vanishes for the
updated model atom. If neglecting 
collisions with hydrogen atoms, the abundance from IR
lines turns out to be, on average, even lower than
that from weak lines. The differences are $\Delta \eps$(IR-
vis)~=~0.05 and 0.13 dex for the model atoms from
Przybilla et al. (2000) and the updated one, respectively.
We found that with updated model atom
and with taking into account collisions with hydrogen atoms,
$\Delta \eps$(IR-vis) = 0, i.e., the abundance averaged
separately over the IR and visible lines is the same,
$\eps$ = 8.74 $\pm$ 0.05. However, there is no agreement
between the 6300 and 6158 \AA\ lines in this case. If
the 3D corrections are applied, then $\Delta \eps$(IR-vis)
= 0.02. In this case
the abundance difference between the visible lines is
reduced, and the rms error becomes smaller
 $\eps = 8.78 \pm 0.03$. 

Table 8 compares the abundances from individual
lines with the results from Caffau et al. (2008) (C08).
The C08(1D) column contains the LTE abundance
derived by Caffau et al. (2008) with the 1D$_{\rm{LHD}}$
model atmosphere. There is agreement between
the abundances from different lines within 0.05 dex,
except for the 6158 \AA\ line for which the difference
is 0.17 dex. A discrepancy is obtained from this
line not only in our paper; the difference of the
abundances from this line between the data from
Caffau et al. (2008) and Asplund et al. (2004) is
0.13 dex. The two C08(HM) columns contain the
non-LTE corrections calculated with the HM74
model atmosphere with and without allowance for
the collisions with hydrogen atoms. The non-LTE
corrections are in agreement within 0.05 dex; the
difference is largest for the strongest 7771~\AA\ line at
\kH~=~0, for which $\Delta_{non-LTE}$ = -- 0.23 dex. Table 9
compares the abundances from individual lines with
the results from Asplund et al. (2004) (A04) obtained
with the MARCS 1D model atmosphere. In the
LTE and non-LTE cases, there is agreement within
0.04 dex for all lines except 6300 \AA , for which the
difference is 0.06 dex. When using the same solar
spectrum, identical atomic data, and the same model
atmosphere (HM74) in Asplund et al. (2004) and
Caffau et al. (2008), the abundance from the forbidden
line differs by 0.09 dex. This is most likely related to
the placement of the continuum.

If the 3D corrections from Caffau et al. (2008) are
added to the non-LTE abundance that we derived
with the updated model, then the line-to-line scatter
is 0.05 dex with \kH~=~1 and
0.9 dex with \kH~=~0. Because of the uncertainties
listed above, it cannot be concluded reliably for
oxygen using only the solar spectrum that \kH~=~1. Therefore, we
performed similar calculations for Procyon (Table 10).
Nobody has performed 3D calculations for O~I
lines in the spectrum of Procyon so far. But the temperature of Procyon is higher, the 3D
corrections are probably smaller than those for the
Sun. In Procyon,
we again divided the lines into two groups, depending
on the non-LTE correction. For weak visible lines,
$\Delta_{non-LTE}$ does not exceed 0.07 dex in absolute value;
for IR lines, $\Delta_{non-LTE}$ > 0.30 dex. The abundances
from the two groups of lines coincide at \kH~=~1, and
the difference between them is 0.13 dex at \kH~=~0.
For Procyon, we can unequivocally conclude that the
collisions with hydrogen atoms with \kH~=~1 should
be taken into account.

For cool stars, where collisions with hydrogen atoms
are more efficient, than collisions with electrons, scaling
the transition rate coefficients under collisions
with electrons does not virtually influence on non-LTE result.
It is worth noting, that for the Sun, when moving from \se~=~1 to \se~=~0.25, 
the non-LTE correction for the 7771 \AA\ line changes only by
0.01 dex. For Procyon, the latter is 0.02 dex.

\begin{figure}  
\resizebox{120mm}{!}{\includegraphics{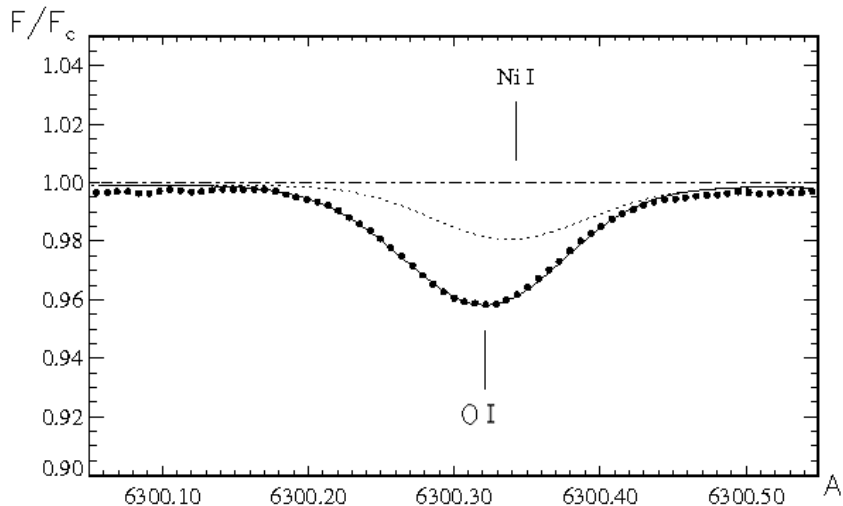}}
\caption{The ${\rm [O~I]}$ 6300~\AA\ line in the solar spectrum. The circles, the dotted line, and the solid line indicate the observed spectrum,
the synthetic spectrum computed without oxygen, and the synthetic spectrum with the oxygen abundance $\eps$ = 8.67 respectively.}
\label{o1_6300}
\end{figure}

\begin{table}  
\caption{ Solar oxygen abundance.}
\label{sun}
\tabcolsep1.0mm
\begin{center}
\begin{tabular}{lrcccccccccc}
\hline\noalign{\smallskip}
$\lambda$ & LTE & \tiny{non-LTE}  & $\tiny{\Delta_{non-LTE}}$  & \tiny{non-LTE} & $\tiny{\Delta_{non-LTE}}$ & \tiny{non-LTE}  & & \tiny{non-LTE}  & $\tiny{\Delta_{non-LTE}}$  & \tiny{non-LTE}  & $\tiny{\Delta_{non-LTE}}$  \\
 \AA&    &   P00 &  P00 & B07  & B07& + 3D  & & P00 &   P00 & B07 & B07     \\
\hline\noalign{\smallskip}
& & \multicolumn{5}{c}{\kH=1}& & \multicolumn{4}{c}{\kH=0}\\
\cline{3-7}
\cline{9-12}
6300 &  8.67 &  8.67 &   0.00 &   8.67 &   0.00 & 8.72 & & 8.67 &    0.00 &  8.67 &   0.00 \\
 6158 &  8.84 &  8.82 & --0.02 &  8.82 & --0.02 & 8.79 & & 8.81 & --0.03 &  8.79 & --0.05 \\
 7771 &  8.92 &  8.78 & --0.14 &  8.74 & --0.18 & 8.80 & & 8.69 & --0.23 &  8.58 & --0.34 \\
 7774 &  8.91 &  8.78 & --0.13 &  8.75 & --0.16 & 8.79 & & 8.70 & --0.21 &  8.59 & --0.32 \\
 7775 &  8.89 &  8.78 & --0.11 &  8.75 & --0.14 & 8.78 & & 8.71 & --0.18 &  8.61 & --0.28 \\
 8446 &  8.86 &  8.76 & --0.10 &  8.74 & --0.12 & 8.77 & & 8.67 & --0.19 &  8.61 & --0.25 \\
\hline
\end{tabular}
\end{center}
\end{table}  %

\begin{table}  
\caption{Comparison of our results with those from Caffau et al. (2008) }
\label{caffau}
\tabcolsep3.0mm
\begin{center}
\begin{tabular}{ccccccc}
\hline\noalign{\smallskip}
$\lambda $ & $\eps$ & $\eps$ &  $\Delta_{non-LTE}$   & $\Delta_{non-LTE}$   & $\Delta_{non-LTE}$  &  $\Delta_{non-LTE}$ \\
\AA & LTE & LTE    & \kH = 1 &  \kH = 1 &    \kH=0 & \kH = 0  \\
 &  & C08(1D) &   & C08(HM)    &        & C08(HM) \\
\hline\noalign{\smallskip}
 6300 &  8.67 & 8.64 &     0.00 &     0.00 &    0.00 &  0.00 \\
 6158 &  8.84  &8.67 &   --0.02 &    0.00 &  --0.03 &  0.00 \\
 7771 &  8.92 & 8.97 &  --0.14 &  --0.16 & --0.23 & --0.28 \\
 7774 &  8.91 &  8.94 & --0.13 &  --0.14 & --0.21 & --0.25 \\
 7775 &  8.89 &  8.92 & --0.11 &  --0.12 & --0.18 & --0.21 \\
 8446 &  8.86  & 8.80 & --0.10 &  --0.08 & --0.19 & --0.15 \\
\hline
\end{tabular}
\end{center}
\end{table}  %

\begin{table}  
\caption{ Comparison of our results with those from Asplund et al. (2004) obtained with the MARCS 1D model atmosphere.}
\label{asplund}
\tabcolsep3.1mm
\begin{center}
\begin{tabular}{ccccccc}
\hline\noalign{\smallskip}
$\lambda$ & LTE & LTE   & non-LTE  & $\Delta_{non-LTE}$ & non-LTE  & $\Delta_{non-LTE}$    \\
\AA &   &  A04 &   &   &  A04  &  A04    \\
\hline\noalign{\smallskip}
 6300 &  8.67 & 8.73 &  8.67 &   0.00 &    8.73 &   0.00 \\
 6158 &  8.84 & 8.80 &  8.81 & --0.03 &    8.77 & --0.03 \\
 7771 &  8.92 & 8.95 &  8.69 & --0.23 &    8.71 & --0.24 \\
 7774 &  8.91 & 8.94 &  8.70 & --0.21 &    8.71 & --0.23 \\
 7775 &  8.89 & 8.91 &  8.71 & --0.18 &    8.71 & --0.20 \\
 8446 &  8.86 & 8.88 &  8.67 & --0.19 &    8.67 & --0.21 \\
\hline
\end{tabular}
\end{center}
\end{table}  %

\begin{table}  
\caption{Oxygen abundance for Procyon.}
\label{procyon}
\tabcolsep3.7mm
\begin{center}
\begin{tabular}{rlcccc}
\hline\noalign{\smallskip}
$\lambda $ &  $\eps_{LTE}$ & $\eps_{non-LTE}$  & $\Delta_{non-LTE}$  & $\eps_{non-LTE}$   & $\Delta_{non-LTE}$  \\
\AA &   &  \kH = 1 &  \kH = 1 & \kH = 0  &  \kH = 0  \\
\hline\noalign{\smallskip}
 6155 &  8.80 &  8.76 & --0.04 &  8.74 & --0.06 \\
 6156 &  8.72 &  8.69 & --0.03 &  8.67 & --0.05 \\
 6158 &  8.79 &  8.74 & --0.05 &  8.72 & --0.07 \\
Mean& 8.77 & 8.73 & & 8.71 &  \\
 7771 &  9.28 &  8.76 & --0.52 &  8.59 & --0.69 \\
 7774 &  9.26 &  8.76 & --0.50 &  8.59 & --0.67 \\
 7775 &  9.18 &  8.73 & --0.45 &  8.57 & --0.61 \\
 8446 &  9.00 &  8.70 & --0.30 &  8.58 & --0.42 \\
 8446 &  9.04 &  8.71 & --0.33 &  8.58 & --0.46 \\
Mean& 9.15 & 8.73 & & 8.58 & \\
\hline
\end{tabular}
\end{center}
\end{table}  %

\section{The non-LTE correction for model atmospheres with different parameters}

We calculated the non-LTE corrections for O~I
lines for Kurucz's classical model atmospheres with
the following parameters: T$_{\rm eff}$ = 5000 - 10000 K at
1000-K steps, log g = 2 (supergiants) and 4 (main-sequence
stars), solar chemical composition, an oxygen
abundance of 8.83, and microturbulence \vt =
2 \kms. The parameter of collisions with hydrogen
atoms is \kH~=~1. The calculations were performed
with the updated model atom with collision rate coefficients
from Barklem (2007) and with rate coefficients
reduced by a factor of 4. The non-LTE correction
and equivalent width for the 7771 \AA\ line are given in
Table 11 and Fig. 6. The non-LTE corrections for the
remaining lines of this multiplet are slightly smaller in
absolute value and behave similarly. Here, the non-LTE correction should be understood as the difference
between the non-LTE and LTE abundances corresponding
to the non-LTE equivalent width. As expected,
the departures from LTE are enhanced with increasing
T$_{\rm eff}$ and decreasing log g, reaching almost two
orders of magnitude in some cases. The non-LTE
corrections for lines with large departures from LTE
are very sensitive to changes in temperature, surface
gravity, and equivalent width (oxygen abundance).
We stress that the calculated non-LTE correction can be
used to derive the non-LTE abundance if not
only the stellar parameters but also the line equivalent
widths are close to the values from Table 11. Whereas
the non-LTE corrections for visible lines in the atmospheres
of main-sequence stars are small (< --
0.05 dex), they cannot be neglected for supergiants
(Table 11). For example, $\Delta_{non-LTE}$ = -- 0.12 dex for
the 6158 \AA\ line at T$_{\rm eff}$ = 7000 K. In the atmospheres of supergiants,
reducing the electron collision rate coefficients by a factor
of 4 affects weakly the SE of O~I,
because collisions are ineffective. The different case is in the
log g = 4 models with T$_{\rm eff}$ > 7000 K. 
Therefore, the curves
in Fig. 6 diverge at T$_{eff}  \simeq$ 7000 K.

To check our non-LTE method for supergiants, we
determine the oxygen abundance for Deneb (Table
12). We do not use the IR lines, because they
are very strong; the observed equivalent width for
the 7771 \AA\ line is 550 m\AA\ (Takeda 1992). The
mean LTE and non-LTE abundances 
from three O~I lines are $\eps$ = 8.76 $\pm$ 0.03 and
$\eps$ = 8.57 $\pm$ 0.01, respectively. For comparison, Schiller
and Przybilla (2008) obtained similar values, $\eps_{LTE}$ =
8.80 $\pm$ 0.07 and $\eps_{non-LTE}$ = 8.62 $\pm$ 0.02. The
small difference in abundance between this stydy and Schiller
and Przybilla (2008)
is most likely because of using a different set of lines.

\begin{table}  
\caption{ Non-LTE corrections for the O~I 7771 \AA\ and 6158 \AA\  lines for a grid of model atmospheres. The equivalent width in m\AA is given in parentheses.}
\label{corr77}
\tabcolsep3.7mm
\begin{center}
\begin{tabular}{rccccccc}
\hline\noalign{\smallskip}
 & \multicolumn{5}{c}{7771 \AA } & &  6158 \AA\\
\cline{2-6}
T$_{eff}, K$ & \multicolumn{2}{c}{log g = 4 } & &  \multicolumn{4}{c}{log g = 2 } \\
\cline{2-3}
\cline{5-8}
        &            B07         &       1/4(B07)   &     &      B07       &    1/4(B07) & & B07, 1/4(B07) \\
 \hline\noalign{\smallskip}
10000 & --1.07 (246)  & --1.26 (272)  & & --1.87 (291)  & --1.96 (302)  & & --0.26 (103) \\
 9000 & --0.95 (260)  & --1.10 (281)  & & --1.64 (301)  & --1.70 (311)  & &--0.18 (120)  \\
 8000 & --0.81 (261)  & --0.90 (275)  & & --1.40 (310)  & --1.46 (318)  & &--0.13 (127) \\
 7000 & --0.65 (221)  & --0.69 (226)  & & --1.28 (296)  & --1.31 (301)  & & --0.12 ( 98)\\
 6000 & --0.34 (133)  & --0.36 (135)  & & --1.07 (233)  & --1.09 (235) & &--0.08 ( 41)  \\
 5000 & --0.10 ( 38)  & --0.10 ( 38)  & & --0.49 (100)  &--0.50 (101)   & & --0.05 (  8)\\
\hline
\end{tabular}
\end{center}
\end{table}  %

\begin{figure}  
\resizebox{190mm}{!}{\includegraphics{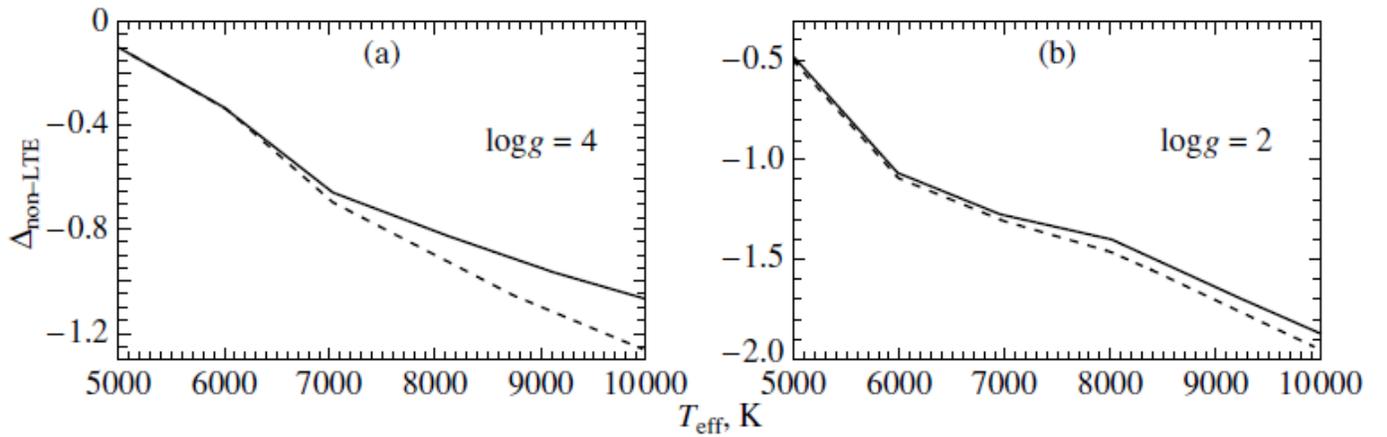}}
\caption{Non-LTE correction for the 7771 \AA\ line versus T$_{\rm eff}$  for (a) main-sequence stars and (b) supergiants. The solid and dashed lines represent the calculations with the original data from Barklem (2007) and with the rate coefficients reduced by a factor of 4, respectively.} 
\label{corr4}
\end{figure}

\begin{table}
\caption{Oxygen abundance for Deneb.}
\label{lines12}
\tabcolsep3.9mm
\begin{center}
\begin{tabular}{rrccc}
\hline\noalign{\smallskip}
$\lambda$, \AA & EW, m \AA& $\eps_{LTE}$ & $\eps_{non-LTE}$  & $\Delta_{non-LTE}$\\
\hline\noalign{\smallskip}
 5330   &  31.4&8.73 & 8.58 & --0.15\\
6155-6 & 117.1& 8.76 &  8.57 & --0.19 \\
 6158  & 92.9 & 8.79 &  8.59 & --0.20 \\
Mean& & 8.76 &  8.57& \\
\hline
\end{tabular}
\end{center}
\end{table}

\section{Conclusions}

We performed non-LTE calculations for O~I with
electron collisional data from Barklem (2007) and determined
the oxygen abundance for six stars.
A significant progress was achieved in reducing the difference
between the abundances derived from IR and
visible lines reduces. It amounts to $\Delta \eps$(IR -vis) = 0.09, 0.14, and
0.14 dex for HD32115, Vega, and Sirius, respectively.
For comparison, $\Delta \eps$(IR -vis)~= ~0.33 dex fov Vega when using the model atom from Przybilla (2000).
To reconcile the abundances from different lines in Vega, we
introduced a scaling factor \se~=~0.25 to the rate coefficients
for collisions with electrons. This value is appropriate for 
other two A-type stars.
The mean abundances, deduced from the visible and IR lines are
$\eps$ = 8.78 $\pm$ 0.09, 8.59 $\pm$ 0.01, and 8.42 $\pm$ 0.03 for
HD 32115, Vega, and Sirius, respectively.

For cool stars (the Sun and Procyon), for which
the main uncertainty of the non-LTE calculations
is associated with allowance for the hydrogen collisions,
we showed that the scatter of the abundances
from different lines is minimal when applying the Drawin's formalism
 in invariable form
(\kH~=~1). The solar oxygen abundance is $\eps$ =
8.74 $\pm$ 0.05 in this case and $\eps_{+3D}$ = 8.78 $\pm$ 0.03
when the 3D corrections from Caffau et al. (2008)
are applied. Scaling the rate of collisions with electrons
affects weakly the abundance; the non-LTE
correction changes by no more than 0.02 dex. The
statistical error in the abundance determined from solar
O~I lines is only 0.03 dex, less than that in Asplund
et al. (2009). However, a conservative estimate of the error
should take into account the change in abundance
due to the use of different observational data ($\sigma$ =
0.01), different continuum placement ($\sigma$ = 0.08), different
model atmospheres ($\sigma$ = 0.06), and different
atomic data ($\sigma$ = 0.02). We estimate the total error to
be 0.11 dex. Note that the solar oxygen abundance we
derived is 0.02 dex higher than that recommended by Caffau
et al. (2008) and 0.09 dex exceeds the value from
Asplund et al. (2009). Nevertheless, it
is lower (by 0.08 dex) than that needed to reconcile
the theoretical and observed density and sound speed
profiles. Recent modeling by Antia and Basu (2011)
showed that not only the oxygen abundance but also
the abundances of heavy elements, such as Ne, are important
to achieve agreement between the theory and
observations. The difference between the theory and
observations is minimized with the chemical composition
of the solar atmosphere that is given by Caffau
et al. (2011) if the neon abundance is increased by a
factor of 1.4.

We calculated the non-LTE corrections for the
O~I 7771 \AA\ and 6158 \AA\ lines for a grid of model
atmospheres with T$_{\rm eff}$ = 5000  -- 10000 K, log g = 2 and 4, solar chemical composition, an oxygen abundance
of 8.83, and microturbulence \vt = 2 \kms.
For IR lines in the atmospheres of supergiants,
$\Delta_{non-LTE}$ reaches almost two orders of magnitude.
The non-LTE corrections for visible lines in supergiants
reach 0.27 dex in absolute value, as distinct
from main-sequence stars in which the non-LTE corrections
for these lines do not exceed 0.05 dex. The calculated corrections
can be applied to determine the non-LTE abundances
if not only the atmospheric parameters but also
the line equivalent widths are close to the tabulated
ones.

\section{Acknowledgments}

We wish to thank D. Shulyak, who computed
the model atmospheres of Vega and HD 32115, and
Yu. Pakhomov for the model atmosphere of Deneb.
We used the DETAIL code provided by K. Butler,
a participant of the "Non-LTE Line Formation for
Trace Elements in Stellar Atmospheres" School
(July 29-August 2, 2007, Nice, France). T.M. Sitnova
and L.I. Mashonkina are grateful to the Swiss
National Science Foundation (the SCOPES project,
IZ73Z0-128180/1) for partial financial support of our
study. The work is supported by a grant on Leading Scientific Schools 
3602.2012.2 and by Federal agency of science and innovations (2012-1.5-12-000-1011-014/8529).

\end{document}